%% file: maxlogapp.tex
\documentclass[journal]{IEEEtran}

\usepackage[cmex10]{amsmath}
\usepackage{pgfplots}
\usepackage{tikz}
\usetikzlibrary{shapes,arrows}
\usepackage{color}
\input{commands}
\begin{document}

\title{Max-log APP Detection\\ for Non-bijective Symbol Constellations}
\author{Martin Damrath, Peter~Adam~Hoeher%
\thanks{M.~Damrath, P.\,A.~Hoeher are with the Faculty of Engineering, 
Kiel University, Kiel, Germany, e-mail: \{md,ph\}@tf.uni-kiel.de.}
}

\maketitle

\begin{abstract}
A posteriori probability (APP) and max-log APP detection is widely used in soft-input soft-output detection.
In contrast to bijective modulation schemes, there are important differences when applying these algorithms to non-bijective symbol constellations.
In this letter the main differences are highlighted.
\end{abstract}

\begin{IEEEkeywords}
Digital modulation, Demodulation, Detection algorithms, Approximation algorithms.
\end{IEEEkeywords}

\section{Introduction}
Commonly used modulation schemes like square quadrature amplitude modulation or phase-shift keying are characterized by a bijective symbol constellation.
Thus, $N$ bits are mapped onto $M=2^N$ symbols.
In contrast, there also exist non-bijective modulation schemes like superposition modulation (SM) \cite{bib:sm_maf}, which typically produce a symbol alphabet with a cardinality less than $2^N$.
This reduction in symbol cardinality can be exploited to achieve power efficiency, bandwidth efficiency, and/or to reduce the detection complexity.
SM is of particular importance, because active signal shaping can be avoided, which makes it naturally near-capacity achieving.
Furthermore the complexity of the optimal detector can be reduced from $\mathcal{O}(2^N)$ down to $\mathcal{O}(N)$ by exploiting the tree-based non-bijective modulation structure.
In the case of non-bijective modulation, redundancy due to channel coding is mandatory.
According to the state-of-the-art, extrinsic information is exchanged between the demodulator (detector) and the channel decoder.
Hence a soft-output detector is mandatory in iterative receivers.
When performing soft-input soft-output (SISO) detection, the a posteriori probability (APP) algorithm provides the optimal solution.
A common simplification of APP detection provides the max-log APP detector \cite{bib:maxlogapp}.
However, these detection algorithms historically have been derived for bijective symbol constellations.
When considering non-bijective modulation schemes, there are important facts to notice when applying APP and max-log APP detection.
Especially for the max-log APP detector, these facts are essential with respect to the performance.
Even if they are rather simple, to the best knowledge of the authors, they have not been published yet.
Furthermore there is no conventional detection method for max-log APP detection of non-bijective modulation schemes in the literature.
In this concise letter the main differences between bijective and non-bijective modulation schemes for implementing the APP detector or max-log APP detector are highlighted, 
and a conventional detection method for the max-log APP detector for non-bijective modulation schemes is proposed.

\section{A Posteriori Probability Detection\label{sec:app}}
For SISO detection, the APP detector provides the optimal solution.
Considering a memoryless channel, the extrinsic log-likelihood ratio (LLR) $L_n$ of bit $b_n$ can be obtained via the well-known formula
\begin{align}
L_n &\doteq \log\frac{p(y|b_n=0)}{p(y|b_n=1)}\\
&=\log\frac{\sum\limits_{\*b_{\sim n}}P(\*b_{\sim n})p(y|\*b_{\sim n},b_n=0)}{\sum\limits_{\*b_{\sim n}}P(\*b_{\sim n})p(y|\*b_{\sim n},b_n=1)}\label{eq:ellr_bit}\\
&=\log\frac{\sum\limits_{x\in\mathcal{X}^{(0)}_n}P(z(x))p(y|x)}{\sum\limits_{x\in\mathcal{X}^{(1)}_n}P(z(x))p(y|x)} \label{eq:ellr_sym} \textnormal{,}
\end{align}
where $y$ represents the channel observation,
$\*b_{\sim n}$ denotes the bit set excluding $b_n$,
$x$ is  a symbol defined over the alphabet $\mathcal{X}$,
$\mathcal{X}_n^{(b)}$ stands for the symbol subset of $\mathcal{X}$ with $b_n = b \in \{0,1\}$,
and $z(x)$ represents the set of all those $\*b_{\sim n}$ that will lead in combination with $b_n=b$ to the symbol $x\in\mathcal{X}_n^{(b)}$.
The terms $P(\*b_{\sim n})$ in (\ref{eq:ellr_bit}) and $P(z(x))$ in (\ref{eq:ellr_sym}) represent a priori information,
and hence are time-varying in the case of iterative processing.
Both (\ref{eq:ellr_bit}) and (\ref{eq:ellr_sym}) are applicable independent of whether the symbol constellation is bijective or non-bijective.
Nevertheless, there is a difference in the number of summands between them for non-bijective symbol constellations.
While each sum in (\ref{eq:ellr_bit}) consists of $2^{N-1}$ summands, each sum in (\ref{eq:ellr_sym}) consists of just $M \leq 2^{N-1}$ summands.

\section{Max-log APP Detection\label{sec:maxlogapp}}
The max-log APP detector \cite{bib:maxlogapp} is a common simplification of the APP algorithm, which avoids the high complexity of exponential and logarithmic operations in APP detection.
Especially in high-SNR scenarios the detector achieves an APP-like performance.
Nevertheless, the complexity still remains to be $\mathcal{O}(2^N)$.
The max-log APP concept is based on the APP detector in the logarithmic domain:
\begin{align}
L_n &= \log\frac{\sum\limits_{\*b_{\sim n}}\e^{\log P(\*b_{\sim n}) + \log p(y|\*b_{\sim n},b_n=0)}}{\sum\limits_{\*b_{\sim n}}\e^{\log P(\*b_{\sim n}) + \log p(y|\*b_{\sim n},b_n=1)}}\label{eq:ellr_log_bit}\\
&= \log\frac{\sum\limits_{x\in\mathcal{X}_n^{(0)}}\e^{\log P(z(x)) + \log p(y|x)}}{\sum\limits_{x\in\mathcal{X}_n^{(1)}}\e^{\log P(z(x)) + \log p(y|x)}} \label{eq:ellr_log_sym}\textnormal{.}
\end{align}
Motivated by 
\begin{align}
\max \phantom{}^*\left(a,b\right) &\doteq \log(e^a+e^b) \\
&\doteq \max\left(a,b\right) + \log\left(1+\e^{-|a-b|}\right) \textnormal{,} \label{eq:correction}
\end{align}
the $\max\phantom{}^*$-operation can be approximated as ${\max\phantom{}^*(a,b)\approx\max(a,b)}$ \cite{bib:maxlogapp,bib:logapp}.
Thus, (\ref{eq:ellr_log_bit}) and (\ref{eq:ellr_log_sym}) can be approximated by:
\begin{align}
\begin{split}
L_n &\approx \max\limits_{\*b_{\sim n}}\left\{\log P(\*b_{\sim n})+\log p(y|\*b_{\sim n},b_n=0)\right\} \\
&- \max\limits_{\*b_{\sim n}}\left\{\log P(\*b_{\sim n}) + \log p(y|\*b_{\sim n},b_n=1)\right\}\label{eq:ellr_bit_ml}\end{split} \\
\begin{split}
L_n &\approx \max\limits_{x\in\mathcal{X}_n^{(0)}}\left\{ \log P(z(x)) + \log p(y|x)\right\} \\
&- \max\limits_{x\in\mathcal{X}_n^{(1)}}\left\{\log P(z(x)) + \log p(y|x)\right\} \label{eq:ellr_sym_ml} \textnormal{.} \end{split}                                                                                                                                                                                                   
\end{align}
While (\ref{eq:ellr_bit_ml}) and (\ref{eq:ellr_sym_ml}) leads to the same results for bijective symbol constellations, 
there is an important difference considering non-bijective symbol constellations.
Due to the fact that there is at least one symbol consisting of more than one bit set $\*b_{\sim n}$,
the results of (\ref{eq:ellr_bit_ml}) and (\ref{eq:ellr_sym_ml}) might no longer be equal.
In fact, (\ref{eq:ellr_sym_ml}) outperforms (\ref{eq:ellr_bit_ml}) and provides a reasonable approximation of the APP algorithm, 
as it will be shown in Section~\ref{sec:numresults}.
Thus, (\ref{eq:ellr_sym_ml}) should be applied as the conventional detection method for non-bijective modulation schemes.
It is also possible to transform (\ref{eq:ellr_bit_ml}) into (\ref{eq:ellr_sym_ml}).
Therefore, all summands corresponding to a bit set $\*b_{\sim n}$, which leads in combination with $b_n$ to the identical symbol $x$, have to be summed up in (\ref{eq:ellr_bit_ml}),
before applying the $\max$-operation:
\begin{equation}
\begin{split}
L_n &\approx \max\limits_{x\in\mathcal{X}_n^{(0)}}\left\{\log\left( \sum\limits_{\*b_{\sim n} \to x} P(\*b_{\sim n})\right) + \log p(y|x)\right\} \\
&- \max\limits_{x\in\mathcal{X}_n^{(1)}}\left\{\log \left( \sum\limits_{\*b_{\sim n} \to x} P(\*b_{\sim n}) \right) + \log p(y|x)\right\} \\
&\approx \max\limits_{x\in\mathcal{X}_n^{(0)}}\left\{\max\limits_{\*b_{\sim n} \to x}\phantom{}^* \left(\log P(\*b_{\sim n})\right) + \log p(y|x)\right\} \\
&- \max\limits_{x\in\mathcal{X}_n^{(1)}}\left\{\max\limits_{\*b_{\sim n} \to x}\phantom{}^* \left(P(\*b_{\sim n}) \right) + \log p(y|x)\right\} \textnormal{.}\label{eq:ellr_bitsym_ml}
\end{split}
\end{equation}
In comparison, (\ref{eq:ellr_bitsym_ml}) is more complex than (\ref{eq:ellr_bit_ml}), 
because the $\max\phantom{}^*$-operator has to be applied.
However, non-bijective symbol constellations often imply a potential for complexity reduction, when exploiting the non-bijectivity in calculation of (\ref{eq:ellr_sym_ml}).
In detail, the construction of $z(x)$ can be visualized as a tree diagram as it is done in \cite{bib:sm_maf}.
Assuming that the complexity of determining $\log P(z(x))$ is proportional to the number of branches in the tree diagram,
the reduction in complexity compared to bijective computation becomes clear.
In the worst case $z(x)$ is bijective and the complexity of both (\ref{eq:ellr_bit_ml}) and (\ref{eq:ellr_sym_ml}) are the same.
The amount of branches in this case is given by $2^N - 2$.
In contrast, if $z(x)$ is non-bijective, the complexity of determining $\log P(z(x))$ can be reduced by reusing several states in the tree diagram for different $\log P(z(x))$.
When applying direct superposition modulation with equal power allocation, the amount of branches needed for $\log P(z(x))$ can be reduced from $2^N - 2$ (for bijective calculation) to $N^2 - N$ (for non-bijective calculation).
Thus (\ref{eq:ellr_sym_ml}) implies a significant potential of complexity reduction compared to (\ref{eq:ellr_bit_ml}).

Given (\ref{eq:ellr_bit_ml}) and (\ref{eq:ellr_sym_ml}), true APP detection is obtained by (re-)substituting all $\max$-operations by $\max\phantom{}^*$-operations.
Hence, the extra computational complexity of APP detection only depends on how to implement the correction term $\log(1+\e^{-|a-b|})$, cf. (\ref{eq:correction}).
Towards this goal, numerous solutions exist, ranging from a simple table look-up \cite{bib:maxlogapp} to the direct implementation of the complex correction term.

\section{Numerical Results\label{sec:numresults}}
\begin{figure}
\begin{center}
\begin{tikzpicture}
\begin{axis}[width=8.5cm, grid=both,
grid style={dotted},
xmin=-2.0, xmax=2.0, ymin=-4, ymax=4,
legend style={legend pos=north west},
legend cell align=left,
xlabel={$\mbox{Re}\{y\}$}, ylabel={$L_n$}]
\addplot [black] table[x=y, y=APP] {./pic/DSM_EPA_LA0.dat};
\addplot [blue, dashed] table[x=y, y=BML] {./pic/DSM_EPA_LA0.dat};
\addplot [red] table[x=y, y=ML] {./pic/DSM_EPA_LA0.dat};
\legend{\small{APP}, \small{Max-Log APP (\ref{eq:ellr_bit_ml})}, \small{Max-Log APP (\ref{eq:ellr_sym_ml})}}
\end{axis}
\end{tikzpicture}
\end{center}
\caption{Relationship between extrinsic log-likelihood ratio $L_n$ and the real part of the channel observation $y$ ($N=16$ DSM-EPA, $\textnormal{SNR}=12\,\textnormal{dB}$).}
\label{fig:yllr}
\end{figure}
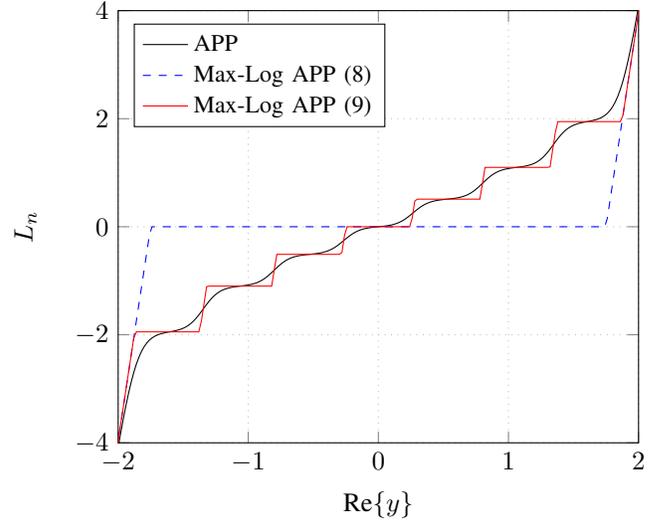
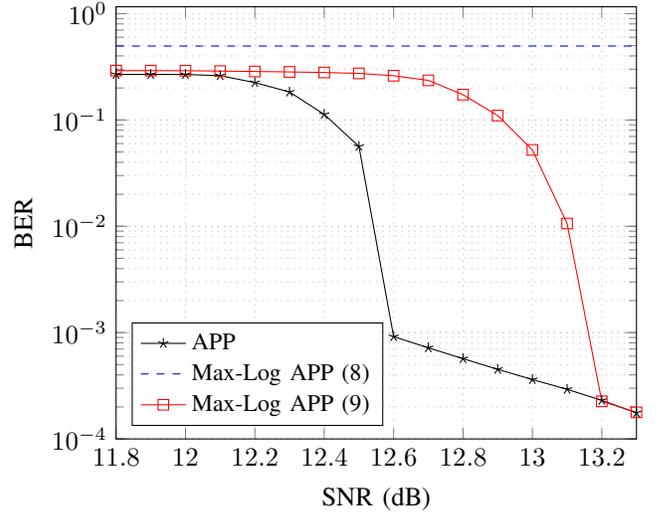
\begin{figure}
\begin{center}
\begin{tikzpicture}
\usetikzlibrary{plotmarks}
\begin{semilogyaxis}[width=8.5cm, grid=both,
grid style={dotted},
xmin=11.8, xmax=13.3, ymin=0.0001,
legend style={legend pos=south west},
legend cell align=left,
xlabel={$\textnormal{SNR}$ ($\textnormal{dB}$)}, ylabel={$\textnormal{BER}$}]
\addplot [black, mark=star, mark repeat=1] table[x=SNR, y=BER] {./pic/app.dat};
\addplot [blue, dashed] table[x=SNR, y=BERsoft] {./pic/main_dsm8_real_maxlogapp.old};
\addplot [red, mark=square, mark repeat=1] table[x=SNR, y=BER] {./pic/maxlogapp.dat};
\legend{\small{APP}, \small{Max-Log APP (\ref{eq:ellr_bit_ml})}, \small{Max-Log APP (\ref{eq:ellr_sym_ml})}}
\end{semilogyaxis}
\end{tikzpicture}
\end{center}
\caption{Bit error rate performance for DSM-EPA ($N=16$).}
\label{fig:ber}
\end{figure}
So far, the derivation is general and not restricted to modulation schemes.
In fact, it can be exploited in many applications like multiple-input multiple-output detection or multi-user detection.
However, the numerical results of this letter are focused on modulation schemes, especially on direct superposition modulation with equal power allocation (DSM-EPA) \cite{bib:sm_maf}.
For the following simulations, the additive white Gaussian noise channel is assumed.
Thus, a channel observation can be written as $y=x+w$,
where $w \sim \mathcal{N}(0,\sigma^2)$ is a zero-mean white Gaussian noise process with the variance $\sigma^2$.
Consequently, the conditional probability is given as
\begin{equation}
p(y|x)=\frac{1}{\pi \sigma^2}\e^{-\frac{|y-x|^2}{\sigma^2}} \textnormal{.} 
\end{equation}

Fig.~\ref{fig:yllr} shows the relationship between the extrinsic log-likelihood ratio and the real part of the channel observations for DSM-EPA with $N=16$ layers for the case that no a priori information is available.
DSM-EPA is strongly non-bijective and leads to $N^2/4+N+1 \ll 2^N$ symbols.
The APP detector from Section~\ref{sec:app} is visualized by the black curve and provides the optimum performance.
The max-log APP detector is realized for the two different implementations discussed in Section~\ref{sec:maxlogapp}: 
Equation (\ref{eq:ellr_bit_ml}) corresponds to the blue curve and (\ref{eq:ellr_sym_ml}) corresponds to the red curve.
As it can be seen from Fig.~\ref{fig:yllr}, (\ref{eq:ellr_sym_ml}) performs a better approximation of the APP detector than (\ref{eq:ellr_bit_ml}) in the case of a non-bijective symbol constellation.
This statement is still true, if a priori information is available.

\begin{table}[!t]
\renewcommand{\arraystretch}{1.3}
\caption{Parameter set of the irregular convolutional code used in the simulation results.
The subcodes are obtained from a recursive systematic convolutional code by puncturing and repetitions, respectively.
The code polynomials and the puncturing/repetition table is given in \cite{bib:ircc}.}
\label{tab:ircc}
\centering
\begin{tabular}{|c||c|c|}
\hline
$j$ & ${R_j}$ & $\alpha_j$\\
\hline\hline
1 & $0.10$ & $0.254042$ \\
2 & $0.15$ & $0.292594$ \\
3 & $0.20$ & $0.003651$ \\
4 & $0.25$ & $0.133594$ \\
5 & $0.30$ & $0.054518$ \\
6 & $0.35$ & $0.032276$ \\
7 & $0.40$ & $0.092666$ \\
8 & $0.45$ & $0.000000$ \\
9 & $0.50$ & $0.000000$ \\
10 & $0.55$ & $0.105838$ \\
11 & $0.60$ & $0.030820$ \\
\hline
\end{tabular}
\end{table}

Concerning bit error rate (BER) simulations, a bit-interleaved coded modulation \cite{bib:bicm} system with iterative processing has been implemented.
In order to achieve a near-capacity performance (at least for APP detection) without active signal shaping, an irregular convolutional code%
\footnote{An irregular convolutional code $C$ of rate $R$ consists of $J$ punctured convolutional codes $C_j$ of rates $R_j$ implemented in parallel,
where $R=\sum\limits_{j=1}^{J}\alpha_j R_j$ and $\sum\limits_{j=1}^{J}\alpha_j=1$.
The EXIT chart characteristic of the irregular convolutional code can be shaped by optimizing the parameter set $\alpha_j$, $1\leq j \leq J$.}
according to \cite{bib:ircc} with a code rate of about $R\approx1/4$ has been matched by means of an EXIT chart design to DSM-EPA employing $N=16$ layers, cf. Table \ref{tab:ircc}.
The bandwidth efficiency is $R\cdot N=4$~bits/symbol, which can theoretically be achieved by $2^{16}$-ary DSM-EPA at an SNR of $12.0\,\textnormal{dB}$. 
The information word length has been chosen to $100\,000$ bits and $500$ iterations between the detector and the channel decoder have been performed.
BER simulation results are visualized in Fig.~\ref{fig:ber}.
The top blue curve shown in broken lines is obtained, when all $\max\phantom{}^*$-operations in an APP detector matched to the symbol alphabet of size $2^N$ are replaced by the $\max$-operation,
which corresponds to (\ref{eq:ellr_bit_ml}). 
Although this is usually exactly what is done quite often in the case of bijective modulation schemes (and referred to as max-log APP detection \cite{bib:maxlogapp}),
it completely fails in the case of non-bijective modulation schemes like DSM: 
The iterative receiver does not converge in the interesting SNR range.
The max-log APP detector according to (\ref{eq:ellr_sym_ml}), which performs the $\max$-operation over the symbol alphabet of cardinality $N^2/4+N+1$, provides a much better performance.
Compared to the APP detector, it degrades only by about $0.56\,\textnormal{dB}$ in the area of the turbo cliff, and the iterative receiver converges.
Thus, (\ref{eq:ellr_sym_ml}) clearly outperforms (\ref{eq:ellr_bit_ml}) in the case of non-bijective symbol constellations.
The error floor shown in Fig.~\ref{fig:ber} can be reduced or avoided by means of doping, which is beyond the scope of this letter, however. 

\section{Conclusion}
It is shown that there are important facts to notice when applying APP and max-log APP detection for non-bijective symbol constellations.
In APP detection, there is no need to distinguish between bijectivity or non-bijectivity, 
but in max-log APP detection, it is important to distinguish between them for achieving the best performance for non-bijective symbol constellations.
The main differences with respect to non-bijective modulation schemes are highlighted and supported by numerical results.
Starting off from an APP detector and replacing all $\max\phantom{}^*$-operations by $\max$-operations, as done quite frequently, completely fails for non-bijective modulation schemes,
thus a conventional detection method for max-log APP detection of non-bijective modulation schemes is proposed.

\bibliographystyle{sty/IEEEtranTCOM}
\bibliography{sty/IEEEabrv,sty/ICTabrv,literatur}

\end{document}

%% file: commands.tex
\def\*#1{\ensuremath{\mathbf{#1}}}

\def\e{\ensuremath{\textnormal{e}}}

\makeatletter
\renewcommand*\env@matrix[1][*\c@MaxMatrixCols c]{%
  \hskip -\arraycolsep
  \let\@ifnextchar\new@ifnextchar
  \array{#1}}
\makeatother